\documentclass[sigconf]{acmart}  

\AtBeginDocument{%
	}

\setcopyright{acmlicensed}
\copyrightyear{2026}
\acmYear{2026}
\acmDOI{XXXXXXX.XXXXXXX}
\acmConference[EASE 2026]{The 30th International Conference on Evaluation 
and Assessment in Software Engineering}{9–12 June, 2026}{Glasgow, 
Scotland, United Kingdom}
\acmISBN{978-1-4503-XXXX-X/2018/06}

\usepackage{hyperref}

\makeatletter
\newcommand{\anchorhere}[1]{\Hy@raisedlink{\hypertarget{#1}{}}}
\makeatother

\usepackage{url}  
\usepackage{xcolor}

\definecolor{quoteClr}{RGB}{ 10, 60, 130} 

\usepackage{tikz}

\usepackage{dirtytalk}

\usepackage{listings}

\usetikzlibrary {positioning}
\usetikzlibrary{shapes.geometric, arrows, backgrounds, fit, calc}

\pgfdeclarelayer{background}
\pgfdeclarelayer{foreground}
\pgfsetlayers{background,main,foreground}

\newcommand{\OCICD}{ICD}
\newcommand{\OCSCD}{SCD}

\newcounter{myexample}
\newenvironment{exmpl}[1][]{\smallskip\refstepcounter{myexample}\par
	\noindent $\blacktriangleright$ \textbf{Example \themyexample:~ #1} \\ 
	\noindent 
	\small
	}{\hfill $\blacktriangleleft$ \par  \smallskip\noindent%
	\ignorespacesafterend \normalsize 
	}

\newcommand{\DpartnerA}[2]{\noindent \textbf{#1:}~\textcolor{magenta}{\say{#2}}}
\newcommand{\DpartnerB}[2]{\noindent \textbf{#1:}~\textcolor{blue}{\say{#2}}}
\newcommand{\DpartnerAC}[2]{\noindent \textbf{#1:}~\textcolor{magenta}{\say{#2}}}
\newcommand{\DpartnerBC}[2]{\noindent \textbf{#1:}~\textcolor{blue}{\say{#2}}}
\newcommand{\DDescription}[1]{\textless *#1*\textgreater}
\newcommand{\CodeFormat}[1]{\mbox{\textit{\textless#1\textgreater}}}
\newcommand{\DCode}[1]{\hfill \CodeFormat{#1}}

\newcommand{\DCodeTwo}[2]{\hfill \CodeFormat{#1} \CodeFormat{#2}}

\newcommand{\DCodeSub}[2]{\hfill \CodeFormat{#1:#2}}
\newcommand{\DDialogOm}{[\ldots]}
\newcommand{\DDialogCutOff}{(!!\ldots!!)}

\newcommand{\Co}[1]{\emph{#1}}
\newcommand{\HDef}{\emph{Definition:~}}
\newcommand{\HRel}{\\\emph{Relationships:~}}

\newcounter{K}
\renewcommand{\theK}{(\arabic{Q})}
\newcommand{\instQ}[1]{\refstepcounter{Q}\anchorhere{#1}\label{#1}\theQ}

\newcounter{Q}
\renewcommand{\theQ}{(\arabic{K})}
\newcommand{\instK}[1]{\refstepcounter{K}\anchorhere{#1}\label{#1}\theK}
\newcommand{\ElemQ}[2]{\renewcommand{\theQ}{#1{\arabic{Q}}}\instQ{#2}}

\newcommand{\Elem}[2]{\renewcommand{\theK}{#1{\arabic{K}}}\instK{#2}}

\newcommand{\newQ}[1]{\ElemQ{Q}{#1}}

\newcommand{\newA}[3]{ %
	(\hyperref[A-#1]{\textbf{\Elem{A}{#1}}} see section \ref{#2}; supported by#3)}
\newcommand{\refA}[1]{\hyperref[#1]{\anchorhere{A-}\label{A-#1}\textbf{\ref{#1}}}}

\newcommand{\Qby}[3]{%
	\textcolor{quoteClr}{\emph{\say{#3}} (\newQ{#2}:#1)}%
}

\usepackage{tabularx,ragged2e}
\usepackage{makecell} 
\newcolumntype{C}{>{\arraybackslash}X} 
\newcolumntype{Q}{>{\Centering\arraybackslash}X} 

\tikzstyle{SuccessFactor} = [circle, minimum width=1.5cm, text width=1.5cm, text centered, inner sep=1pt, draw=black, fill=green!30, font=\scriptsize]
\tikzstyle{RiskFactor} = [circle, minimum width=1.cm, text width=1.5cm, text centered, inner sep=1pt, draw=black, fill=red!30, font=\scriptsize]
\tikzstyle{Task} = [rectangle, minimum width=1.8cm, minimum height=1cm, text width=1.7cm, inner sep=1pt, text centered, draw=black, fill=gray!10, font=\scriptsize]
\tikzstyle{SubTask} = [rectangle, minimum width=1.6cm, minimum height=1cm, text width=1.5cm, inner sep=1pt, text centered, draw=black, fill=gray!5, font=\scriptsize, dashed]
\tikzstyle{BadTask} = [rectangle, minimum width=1.8cm, minimum height=1cm, text width=1.7cm, inner sep=1pt, text centered, draw=black, fill=red!10, font=\scriptsize]
\tikzstyle{BadSubTask} = [rectangle, minimum width=1.6cm, minimum height=1cm, text width=1.5cm, inner sep=1pt, text centered, draw=black, fill=red!5, font=\scriptsize, dashed]
\tikzstyle{Legend} = [rectangle, minimum width=2cm, minimum height=1.5cm, text width=2cm, inner sep=4pt, text centered, draw=black, font=\scriptsize]
\tikzstyle{posArrowL} = [green!50!black!50, , thick, line width=1mm,->,>=stealth]
\tikzstyle{posArrowS} = [green!50!black!50, thick, line width=0.66mm,->,>=stealth]
\tikzstyle{negArrowL} = [red!50!black!60, thick, line width=1mm,->,>=stealth]
\tikzstyle{negArrowS} = [red!50!black!60, thick, line width=0.66mm,->,>=stealth]

\tikzstyle{subConceptA} = [o-stealth, black, thick, dashed]

\usetikzlibrary{patterns}

\newcommand{\basicTimeline}[1]{
	\newcounter{recHour}
	\setcounter{recHour}{0}
	\newcount\minuteOne; \minuteOne=0 
	\def\w{0.8\columnwidth}    
	\def\n{#1}     
	\def\lt{0.40} 
	\def\lf{0.36} 
	\def\lo{0.30} 

	\draw[->,thick] (-\w*0.03,0) -- (\w*1.03,0);
	
	\foreach \tick in {0,1,...,\n}{
		\def\x{{\tick*\w/\n}}
		\def\minute{\the\numexpr \minuteOne+\tick*10-\value{recHour}*60 \relax}
		\ifnum\minute > 59
		\stepcounter{recHour}
		\def\minute{\the\numexpr \minuteOne+\tick*10-\value{recHour}*60 \relax}
		\fi
		\draw[thick] (\x,\lt) -- (\x,-\lt) 
		node[below] {
			\ifnum\minute<10
			{\footnotesize \the\value{recHour}:0\minute h}
			\else
			{\footnotesize \the\value{recHour}:\minute h}
			\fi
		};
		
		\ifnum \tick<\n
		\draw[thick] ({(\x+\w/\n/2)},0) -- ({(\x+\w/\n/2)},\lf); 
		\foreach \ticko in {1,2,3,4,6,7,8,9}{
			\def\xo{{(\x+\ticko*\w/\n/10)}}
			\draw[thick] (\xo,0) -- (\xo,\lo);  
		}\fi
	}
}

\def\minuteLabel(#1,#2){
	\node[above] at ({(#1-\minuteOne)*\w/\n/10},\lt) {#2};}

\def\minuteArrowLabel(#1,#2,#3,#4){ 
	\def\xy{{(#1-\minuteOne)*\w/\n/10}}; \pgfmathparse{int(#2*100)};
	\ifnum \pgfmathresult<0
	\def\yyp{{(\lt*(0.90+#2))}}; \def\yyw{{(\yyp-\lt*#3)}}
	\draw[<-,thick,black,align=center] (\xy,\yyp) -- (\xy,\yyw) node[below,black] at (\xy,\yyw) {#4};
	\else
	\def\yyp{{(\lt*(0.10+#2)}}; \def\yyw{{(\yyp+\lt*#3)}}
	\draw[<-,thick,black,align=center] (\xy,\yyp) -- (\xy,\yyw) node[above,black] at (\xy,\yyw) {#4};
	\fi}

\begin{document}
	
\title[Managing Power Gaps as an Element of 
	Pair Programming Skill:	\\ A Grounded Theory]{Managing Power Gaps as an Element of 
	\\Pair Programming Skill:
	A Grounded Theory}

\author{Linus Ververs}
\email{linus.ververs@fu-berlin.de}
\affiliation{%
	\institution{Freie Universität Berlin}
	\country{Germany}
}

\author{Janina Berger}
\email{janib03@zedat.fu-berlin.de}
\affiliation{%
	\institution{Freie Universität Berlin}
	\country{Germany}
}

\author{Lutz Prechelt}
\email{prechelt@inf.fu-berlin.de}
\affiliation{%
	\institution{Freie Universität Berlin}
	\country{Germany}
}

\renewcommand{\shortauthors}{Ververs, Berger \& Prechelt}

\begin{abstract}
  \emph{Background:} In pair programming, Togetherness (the partners
    understand each other's mental state well) is a main success factor.
    Maintaining high Togetherness is an element of pair programming skill.
    Some sessions appear to go badly although Togetherness appears good.
	\\
  \emph{Objective:} Understand under what circumstances this is possible.
	\\  
  \emph{Method:}
	Grounded Theory Methodology based on
    21 recorded pair programming sessions with 22 developers
	from 5 German software companies and
	6 interviews with different developers from 4 other German companies.\\
  \emph{Results:} We explain how a Power Gap can make a session dysfunctional
    despite the presence of high Togetherness, how it comes into existence
    due to a Knowledge Gap and Hierarchical Behavior,
    why its consequences (Defensive Behavior and Disengaging Behavior)
    are problematic, and how it can be reduced or prevented by
    Equalizing Behavior.
	\\
  \emph{Conclusions:}
    Pair programming practitioners can improve their pair programming skill
    by unlearning problematic behaviors related to Power Gaps
    and by learning to recognize Power Gaps and apply Equalizing Behavior.
\end{abstract}




\maketitle
	

\section{Introduction}

In the most-cited definition pair programming is defined as \say{two 
programmers jointly produce one artifact (design, algorithm, code)} 
\cite{WilKesCun00}. 
Pair programming is used widely across the industry \cite{BegNag08, StackOverflow2018, DigitalStateofAgile2020}.
Most often, pair programming is not used as a practice, i.e.,
for all production code without exception \cite{Beck99,ComSilSuc08}, but rather
mostly in development situations where the
task is particularly difficult \cite{HanDybAri09} because of
the advantages one can expect from using it:
knowledge sharing and transfer between participating developers \cite{PloShaLin15,Zie20,ZiePre20,VanLas05, VanMan13}, 
better design and code quality
\cite{HanDybAri09, FuGraBro17, VanLas05, Zac11, AriGalDyb07, VanMan13}, and 
increased speed \cite{HanDybAri09, SilSucVla12} 
(reduced task duration appears to be the least likely of these \cite{AriGalDyb07, Zac11}).
Thus, doing pair programming well is an important success factor
for software development.

We are interested in understanding how to do pair programming well.
Previous research found this to be a question of pair programming skill,
a capability that is separable from general software development skill \cite{ZiePre21}.
The key element of that skill appears to be the habit of maintaining high
Togetherness, which is defined as
``the degree to which the pair members are able to fully understand each other’s
activities, including all intentions and meanings associated by the 
respective speaker or actor.'' \cite[Section 6.4.1]{Zie20}.
The only other elements that have been formulated so far have to do
with avoiding problematic behaviors:
make sure to end your knowledge transfer episodes properly,
do not drown your partner in exaggerated explanations \cite{ZiePre21}.

Our research looks for further elements of pair programming skill.
Our starting point were cases of pair programming behavior 
that felt dysfunctional even though the Togetherness appeared to be alright, 
so we attempted to understand what was going on in them.
Here, we aim at providing a scientific analysis of the 
mechanisms underlying many pair programming dysfunctionalities.

\subsection{Research Interest}

As we use Grounded Theory Methodology, we do not formulate fixed, inflexible
research questions, but rather a less specific \emph{research interest}.
Our research interest is understanding how pair programming may become
dysfunctional \emph{despite} apparently good Togetherness and
which elements of pair programming skill might help avoid this.

\subsection{Research Contributions}

We identify a risk factor (\Co{Power Gap}) that threatens all major
advantages of pair programming.
We identify its antecedent, \Co{Hierarchical Behavior}, 
a behavior that will often produce a \Co{Power Gap}.
We identify frequent consequences, \Co{Defensive Behavior} and 
\Co{Disengaging Behavior}, that tend to break the collaboration in a 
pair programming team.
We explain a compensating behavior, \Co{Equalizing Behavior},
with which pair programmers can keep a \Co{Power Gap} in check and 
preserve their collaboration.

These behaviors (three of them problematic, one helpful)
extend our understanding of pair programming skill.

\subsection{Article Overview}  

We will proceed to describe
our data
(Section~\ref{SecPPSessions}),
our coding process 
(Section~\ref{SecCoding}), and
the manner in which we define concepts
(Section~\ref{SecConceptDefs}).
We continue by defining our main concepts
(Sections~\ref{sec:KGap} to \ref{sec:ProcessAgency}),
putting the pieces together in 
our grounded theory
(Section~\ref{sec:GT})
and applying it to our notion 
of pair programming skill
(Section~\ref{sec:ppskill}).
We discuss limitations
(Section~\ref{sec_lim}) and
related work
(Section~\ref{SecRelatedWork}),
before we conclude with
implications for practitioners
and further work
(Section~\ref{sec_conclusion}).

\section{Methods and Data} \label{sec:methods}

We use Grounded Theory Methodology (GTM) mostly based on Strauss \&
Corbin's interpretation~\cite{StrCor90}, but also incorporate Charmaz's
constructivist perspective and warnings that
\say{researchers are part of what they study, not separate from it}
\cite{charmaz06,charmaz14}.


For our research, we rely on the PP-ind repository, a preexisting data set 
of approximately 100 hours of recorded pair programming sessions from 13 
companies collected between 2006 and 2018~\cite{ziepre20-pp-ind}. The 57 
participating developers worked on real tasks in their company software; 
the videos contain IDE screen capture plus video/audio, and all developers 
consented to the recordings being used for research.

When our most central concept (\Co{Power Gap}) turned out to be not directly observable,
we led six developer interviews (I1-I6) in four different companies 
as Theoretical Sampling for grounding the concept.
Our university does not require ERB approval for
such interviews with peers.
All transcripts of our interviews (in German) are available
on Figshare\footnote{\url{https://figshare.com/s/24b1e48102189edd6853}}.
The video data contains company secrets and allows identifying developers;
it can therefore not be made available publicly.

\subsection{Session Description} \label{SecPPSessions}
The Grounded Theory (GT) presented in this paper is derived from a total of 21 recorded sessions.
These sessions follow the naming convention of the PP-ind repository~\cite{ziepre20-pp-ind},
where the first letter indicates the company, the second letter denotes the technical context (the software system the pair was working on), and the number specifies the chronological order of pair programming sessions within that context.
For instance, sessions DA2 and DA4 were conducted in company D, both within context A, with DA2 preceding DA4 in time. 

Developers are labeled according to their company and the chronological order of their first appearance in the sessions.
For instance, D1 refers to the first developer from company D who appears in the dataset.

We analyzed sessions AA1, CA1, CA2, CA5, DA2, DA4, DA5, DA6, JA1-JA5, and PA1-PA4. 
We will present and discuss episodes taken from AA1, CA1, CA5, DA2, DA4, DA5, and PA3.
We will introduce the session context along with the episodes as needed.

\subsection{Open Coding \& Code Development} \label{SecCoding}
Initially and for the majority of our analysis we worked with 
the video recordings of the sessions. 
In the early stages of our Open Coding we decided to apply the 
practice of defining (and constantly refining) a \say{perspective on the 
data} as suggested by \citeauthor{SalPloPre08} which acts as a filter to 
only introduce new concepts that are expected to drive the analysis 
forward~\cite{SalPloPre08}.
This is contrary to the suggestions by \citeauthor{StrCor90} who describe 
Open Coding as \say{taking apart an observation, a sentence, a paragraph, 
and giving each discrete incident, idea, or event, a name, something that 
stands for or represents a phenomenon}~\cite[p. 63]{StrCor90}.
We did so because staying open to all relevant concepts would quickly lead 
to us drowning in the data as described by 
\citeauthor{SalPloPre08}~\cite{SalPloPre08} and which we initially 
experienced when we started our Open Coding due to the richness of our 
observational data.  
Additionally, we highly utilized memo writing which drove our analysis 
forward. 
When we encountered an episode in our data that seemed to be relevant, 
i.e., that we perceived to be dysfunctional in some way, we wrote a memo 
about it. 
Later, we revised these memos and introduced codes to investigate and 
break down what actually happened during that episode.    

Here, we explain how this coding process and the defined perspective on 
the data evolved, and which concepts emerged when,
providing detail only for concepts that are not relevant anymore
in the final results and pointing forwards for the rest.

\subsubsection{First Generation of Codes: Conversational Defects}
We have long been interested in things that go badly during pair programming.
Our initial open coding showed these dysfunctionalities as \say{hiccups} in the pair's dialog. 

To investigate this further we applied Zieris' concepts on process
fluency, where utterances are classified into five different base
activities \cite[Section 6.2.2]{Zie20}.
The first four model the \say{conversational role},
the fifth is the \say{conversational defect}:
\say{A missed connection in the dialog [\ldots]; can be a non-action}.

Zieris uses these codes to draw conclusions about process fluency, i.e., 
to identify focus phases (fast, smooth process) and breakdowns, 
the poles of the process fluency spectrum. 
Fluency allows inferring Togetherness of the pair, which is Zieris' key concept.

We found that Conversational Defect utterances not only happen if the pair's Togetherness is low but also in other cases. 
Apparently, Zieris' theory did not explain all dysfunctionalities. 
We wanted to explain their cause and how to prevent them. 
It turned out that accountability issues were productive for
differentiating types of Conversational Defect. 

\emph{Situational Conversational Defects} (\OCSCD)
are caused by the evolving situation, e.g., 
a misunderstanding or when one partner is missing knowledge. 
No one is directly responsible. 
These happen occasionally and are barely problematic. 

\emph{Intentional Conversational Defects} (\OCICD)
are directly caused by partner P's unwillingness to cooperate or engage. 
A root cause for this behavior (e.g. impatience) may lie elsewhere in
the session but right then P \emph{could} clearly react in a
constructive manner.
Rather, P ignores the partner completely (a non-reaction), 
ignores the content of the partner's last utterance, or
interrupts the partner with something unrelated.

ICDs capture dysfunctional behavior not explained by Togetherness.

\subsubsection{Second Generation of Codes: Root Cause Analysis}
After understanding what went wrong by defining ICDs, our next goal was to understand why they occurred.
To achieve this goal, we first needed to understand the context. 

Salinger and Prechelt created the \say{Base Layer} which is a collection of concepts that serve as 
\say{a foundation for qualitative research into pair programming [...] that aims at explaining the pair programming	process} \cite{SalPre13}. 
These concepts provide the context we needed by describing for instance whether
an ICD happens as part of a knowledge transfer episode, or as part of a decision-making episode.
We searched for commonalities of the different ICD contexts,
but found none; ICDs appeared in a wide variety of situations. 

Next we investigated the Driver/Observer-Dynamic. 
Who was the driver seemed to make a crucial difference in how these situations played out, so we annotated utterances with the speaker's role and later also
with a mode of doing, e.g., propose\_design was refined into driver-propose\_design, which was refined into driver-propose\_design-by\_doing, driver-propose\_design-by\_verbalizing\_\-before, and driver-propose\_design-by\_doing\_and\_verbalizing. 

We also coded various additional aspects such as interrupting the partner, 
raising the voice, showing signs of annoyance, disengaging etc.
This traced why individual pairs behaved in the way they did, 
but every pair appeared unique and we were unable to find unifying themes 
and behavioral patterns. 
Our research came to a standstill.

\subsubsection{Third Generation of Codes: Identifying our Core Concepts}

As is typical for Grounded Theory work, the breakthrough occurred quickly
once a single key idea was identified.
In our case, it was the insight that ICDs were produced
by one pair member because they were fed up with how their
partner had behaved before.

Following this trail quickly led 
to classifying the ICDs as \Co{Defensive Behavior}, 
to assuming the reason for it to be the perception of a \Co{Power Gap},
and therefore to classifying the partner's previous behavior as
\Co{Hierarchical Behavior}.
All three will be explained and illustrated in detail in 
Section~\ref{sec:res}.

We found the other concepts of our eventual Grounded Theory
in due course, because the above three served as a powerful
theoretical lens for directing our attention.

\subsection{Manners of Concept Definition}\label{SecConceptDefs}

The results section will present many GTM concepts.
Most of these are new, but a few have been introduced in earlier
works, which we then cite.
Of the new ones, those that will directly be part of the Grounded
Theory (GT concepts) will be defined by an entire section.
Others are auxiliary and are defined much more compactly.
Each GT concept can in principle be subdivided into many subconcepts.
In a few cases, we do this explicitly, but only to make the superconcept's
definition easier to understand.


\section{Results} \label{sec:res}

In this section, we will describe the three main concepts mentioned
just above:
\Co{Defensive Behavior} (in Section~\ref{sec:DefBehav}),
\Co{Power Gap} (Section~\ref{sec:PGap}), and
\Co{Hierarchical Behavior} (Section~\ref{sec:HBehav}).
Each subsection provides a concept definition, 
provides plenty of examples for illustrating the phenomenon in question, and
explains the relationships to other concepts.

The other subsections \ref{sec:KGap} and \ref{sec:DisBehav} to \ref{sec:ProcessAgency}
introduce further main concepts appearing in our Grounded Theory,
culminating in the Grounded Theory itself (Section~\ref{sec:GT}),
which we then relate back to pair programming skills 
(Section~\ref{sec:ppskill}).

Figure~\ref{fig_GT} in Section~\ref{sec:GT} summarizes our Grounded Theory. 
It may help the reader by understanding the relationships between the 
concepts while working through the subsections.

\subsection{Knowledge Gap} \label{sec:KGap}

\HDef We say partner A has a \Co{Knowledge Gap} with respect to partner B if
B possesses task-relevant knowledge that A is currently lacking.
A  \Co{Knowledge Gap} always concerns substantial amounts of knowledge;
small amounts of knowledge that can be transferred in a few seconds
do not constitute a \Co{Knowledge Gap}.
\HRel \Co{Knowledge Gaps} are one of the major drivers of \Co{Power Gaps}.

A common reason for doing pair programming in the first place is when
both partners have a \Co{Knowledge Gap}.
\Co{Knowledge Gaps} usually shrink during a session.

As the \Co{Knowledge Gap} represents the delta between the partner's 
standard of knowledge regarding the current task, we annotate it based on
markers such as asking questions, verbal and non-verbal signs of confusion, 
the other partner explaining proactively, and direct statements regarding
one's own standard of knowledge. 

It is normal and expected to see some of these markers in a session
because knowledge transfer is a key aspect of pair programming. 
For our theory this \say{normal} knowledge transfer is not particularly
relevant. 
Rather, we are interested in cases where it becomes one-sided.
For this reason, we only assign the concept \Co{Knowledge Gap} to a
chunk of a session if we see more markers than usual and
if they stem almost exclusively from one partner. 

Example~\ref{DA2_PropSwitch} illustrates the effect of a \Co{Knowledge Gap}.

\begin{exmpl}[DA2 - Proposed Driver/Observer Switch] \label{DA2_PropSwitch}
	It is D4's first week in company D. 
	D3 is tasked with onboarding D4.
    They implement a new UI feature in Java. 
	Their interaction quickly changes as newcomer D4 has more
    general knowledge of Java and software design than D3,
    and D4 soon becomes the leading partner. 
	We see longer sections of \Co{Unbalanced Process Agency}
    (see Section \ref{sec:ProcessAgency}) as a result. 
	Just before the following episode, the pair had received help from
    a third developer.
	
	\DpartnerAC{D4}{Uh, do you want to take over again?}
	
	\DpartnerBC{D3}{I think you're more involved in the whole thing.
      I'm\ldots for me, 
	  I am working at the limits of my knowledge. \DDescription{Laughing.}} 
	
	\DCodeTwo{Disengaging Behavior}{knowledge gap marker} 
	
	\DpartnerA{D4}{\DDescription{Laughing.}}

    We had seen several markers of a \Co{Knowledge Gap} before this episode
    and now D3 even states it explicitly when refusing the driver role.
\end{exmpl} 

We will refer to this example a second time when we discuss
\Co{Disengaging Behavior} in Section~\ref{sec:DisBehav}.

\subsection{Defensive Behavior} \label{sec:DefBehav}

\HDef \Co{Defensive Behavior} is passive-aggressive behavior
where a developer minimizes communication by shutting out their partner.
\HRel It is a subordinate partner's reaction to a big 
\Co{Power Gap}, which has appeared due to the
other's previous \Co{Hierarchical Behavior}.

Example~\ref{PA3_AssertSame} shows \Co{Defensive Behavior} through \Co{Minimizing Communication}:

\begin{exmpl}[PA3 - Minimizing Communication] \label{PA3_AssertSame}
	P1 and P3 currently work on a test case. It compares the result of a 
	function to a predefined array. The pair is uncertain which 
	assert-function to use.
	
	\DpartnerBC{P1}{So now, of course, with \texttt{assertSame} you have the additional condition that the two share the same memory area, that they have the same memory} 
	
	\DCodeSub{Hierarchical Behavior}{Lecturing}
	
	\DpartnerA{P3}{\DDescription{P3 searches for an alternative to \texttt{assertSame} using the IDE suggestions while P1 speaks.} That this is the identical object.} 
	
	\DpartnerB{P1}{Exactly. Um.}
	
	\DpartnerA{P3}{\DDescription{scrolls to \texttt{assertEquals}} We want equals I suppose.} 
	
	\DpartnerB{P1}{Equals would be \DDialogCutOff\footnote{We use this notation if the current speaker is interrupted by their partner.}}
	
	\DpartnerA{P3}{I don't know if this works with Arrays.} 
		
	\DpartnerBC{P1}{Equals is better, you can also use it to compare arrays. Mmh in case of doubt, you have a dependency on the order of the individual items that they contain, which is also \DDialogCutOff} 
	\DCodeSub{Hierarchical Behavior}{Lecturing}
	
	\DpartnerAC{P3}{There is even \texttt{maxDepth} \DDescription{Points to the screen.}} 
	
	\DCodeSub{Defensive Behavior}{Minimizing Communication} 
	
	\DpartnerB{P1}{Not necessarily.}
	
	\DpartnerA{P3}{So cool!}
	
	\DpartnerBC{P1}{Er, which is not necessarily that important. But if you still want to ensure the data types, in other words you want to say, okay this is really a string that is returned, this is really a float \DDialogCutOff} 
	
	\DCodeSub{Hierarchical Behavior}{Lecturing}
	
	\DpartnerAC{P3}{Mhm. \DDescription{continues scrolling through the IDE suggestions.}} 
	
	\DCodeSub{Defensive Behavior}{Minimizing Communication} 
	
	\DpartnerB{P1}{or an object \DDialogCutOff}
	
	\DpartnerAC{P3}{Look, there is \DDialogCutOff}
	
	\DCodeSub{Defensive Behavior}{Minimizing Communication} 
	
	\DpartnerBC{P1}{You can also iterate.} 
	\DCodeSub{Hierarchical Behavior}{Lecturing}
	
	\DpartnerA{P3}{There is also json.}
	
	\DpartnerBC{P1}{You can also use that.} 
	\DCodeSub{Hierarchical Behavior}{Lecturing}
	
	\DpartnerA{P3}{\DDescription{Continues scrolling through the IDE suggestions.}}
	
	\DpartnerBC{P1}{There are quite a lot of functions that I... I have never really made use of during normal implementation. Sometimes \DDialogCutOff} 
	
	\DpartnerAC{P3}{So what did you say?}
	
	\DpartnerB{P1}{PhpStorm recommends something.}
	
	\DpartnerA{P3}{What's better than Equals? What did you just want to say?} 
	
	\DpartnerB{P1}{So \texttt{assertEquals} would be one option}
	
	\DpartnerA{P3}{\DDescription{P3 selects \texttt{assertEquals} from the IDE suggestions.}}
	
	\DpartnerBC{P1}{But then of course you would have to use... uhh \texttt{assertSame}. Sorry \texttt{assertSame}. But then you just iterate over all the values of the array. So you do a foreach with an \texttt{assertSame}} 	
	\DCodeSub{Hierarchical Behavior}{Lecturing}
	
	\DpartnerA{P3}{Oh well.}
	
	\DpartnerB{P1}{This ensures that the key you expect is defined and that the data type you have there is really the same and not just uh convertible to this data type.}	
	\DCodeSub{Hierarchical Behavior}{Lecturing}
	
	\DpartnerAC{P3}{Let's do Equals first... I would say. \DDescription{P3 implements a naive check ignoring P1's proposal on looping over the elements.}} 
	
	\DCodeSub{Disengaging Behavior}{Hiding Knowledge Gap}
	
	\DpartnerBC{P1}{In this case, you are also responsible for ensuring that the order is correct. This is always nasty with array comparisons.} 
	
	\DCodeSub{Hierarchical Behavior}{Lecturing}
	
	\DpartnerAC{P3}{So.\DDescription{Talking more to himself and signaling he is done with this decision}}
	\DCodeSub{Defensive Behavior}{Minimizing Communication} 
		
	We will refer to this example again when talking about 
	\Co{Hierarchical Behavior} in Section~\ref{sec:HBehav}. 
	Ignore the \Co{Hierarchical Behavior} annotations for now.
	In this example, we see four cases of 
	\Co{Minimizing Communication}.\footnote{Essentially what we 
		called {Intentional Conversational Defect} 
		in the early phase of our analysis.} 
	P3 demonstrates this by interrupting P1, starting new topics,
	or not engaging	content-wise with P1. 
	Consequently, the pair decides to use \texttt{AssertEquals}, 
	even though \texttt{AssertSame} would have been the more appropriate choice.
\end{exmpl}

\Co{Defensive Behavior}, by reducing communication, will usually damage
the \Co{Togetherness} of a pair.
Lower \Co{Togetherness} increases the risk of process breakdowns~\cite[Chapter 6]{Zie20},
which hurt the efficiency of the process.

\subsection{The Power Gap} \label{sec:PGap}

\HDef A \Co{Power Gap} is the perception by one partner of the pair
that the opportunities for participating in the pair process are not
distributed equally.
We call the partner with fewer perceived opportunities \Co{subordinate},
the other \Co{dominant}.
\Co{Power Gaps} are not static; they may change during the session.
Sometimes, both partners can perceive themselves to be at the subordinate end
of a \Co{Power Gap}.
\HRel The most common cause and trigger of a \Co{Power Gap} is a
\Co{Knowledge Gap}.
Its most prominent effects are
\Co{Defensive Behavior} and/or \Co{Disengaging Behavior}.

The \Co{Power Gap} is subjective and rarely addressed explicitly by the
pair members, which means the grounding will have to involve related concepts.

For instance, the reason for P3's \Co{Minimizing Communication} behavior 
in Example~\ref{PA3_AssertSame}
was stress he had endured during most of the session,
which resulted from a \Co{Power Gap}, which resulted from a \Co{Knowledge Gap}
and P1's \Co{Hierarchical Behavior}.
We have not explained \Co{Hierarchical Behavior} yet,
but the point of the above statement is a different one:
To understand even this single instance of \Co{Power Gap},
we first need to understand several related concepts
(like \Co{Hierarchical Behavior}).
We therefore postpone example-based discussion of \Co{Power Gap} until
the relevant related concepts have been introduced.

\subsection{Hierarchical Behavior} \label{sec:HBehav}

\HDef \Co{Hierarchical Behavior} is any utterance or action where
the active part is perceived by the partner to have a more powerful position. 
(A more precise name for it would be
\Co{Behavior That is Perceived as Hierarchical}. We use the above for brevity.)
\HRel \Co{Hierarchical Behavior} creates or increases a \Co{Power Gap}.

Because it involves a perception, this concept, like the \Co{Power Gap},
is grounded indirectly:
When we encounter a behavior that appears to be hierarchical
(based on markers such as tone of voice, demeaning judgments,
seizing control of the keyboard etc.),
we examine how the partner reacts to the behavior.
If the partner acts coolly, like business-as-usual,  
we consider the behavior not to be \Co{Hierarchical Behavior}. 
Examples~\ref{CA1_NotHier} and~\ref{AA1_MiniObject} show
non-hierarchical and hierarchical behavior, respectively.

\begin{exmpl}[CA1 - Non-Hierarchical Rushing Ahead] \label{CA1_NotHier}
	C1 started working on the task prior to the pair programming session
	with C2. C2 quickly takes over and proposes a different approach.
	
	\DpartnerB{C2}{Oh God\ldots You know what, let’s do this differently.
		\DDescription{Starts editing the code.}}
	
	\DpartnerA{C1}{How?} 
	
	\DpartnerB{C2}{Let’s leave it like this for now. 
		What I’d like is a method that\ldots lets us keep track of 
		those things in the list and then just pass the list. \DDialogOm~ 
		So basically, an add method.
		\DDescription{Begins implementing the method.}}
	
	\DpartnerA{C1}{Okay, I think that’s a bit cumbersome, but alright.} 
	
	\DpartnerB{C2}{\DDescription{No reaction. Continues editing the code.} 
		There, something like that.}
	
	C2 does not wait for an approval to his approach by C1. 
	C1 reacts calmly although he disagrees with the new approach.
	C1 voices his disagreement directly. 
	This is why it is not \Co{Falling in Line} which we will discuss later.
	As we see no signs that C1 interpreted C2 as hierarchical, we do not
	code it as such.
\end{exmpl}

\begin{exmpl}[AA1 - Hierarchical Rushing Ahead] \label{AA1_MiniObject}
	A2 is driving the whole session. 
	They just finished implementing a new method where an object 
	from the class \texttt{MicroObject} is initialized. 
	This exchange starts with A1, who presumably voices 
	a new proposal but is cut off by A2.  
	
	\DpartnerA{A1}{Can you \DDialogCutOff}
	
	\DpartnerBC{A2}{Now we’ll just make a proper 
		\texttt{MiniObject} out of this.
		\DDescription{A2 changes the class name from 
			\texttt{MicroObject} to 
			\texttt{MiniObject}.}} 
		
		\DCode{Hierarchical Behavior}
	
	\DpartnerA{A1}{Nooo! \DDescription{Aggressive tone and shouting.}}
	
	\DpartnerB{A2}{Yes, of course I have to!
		\DDescription{annoyed tone}} 	
	
	Similar to C2’s behavior in Example~\ref{CA1_NotHier}, 
	A2 rushes ahead and does not wait for A1’s approval.
	In contrast to the previous example, 
	A1 does not remain calm but reacts in an aggressive manner. 
	He seemingly interprets A2’s action as hierarchical, 
	i.e., going over his head.
	Otherwise, we see no reason why A1 would react so strongly 
	to such a minor and easily correctable mistake.
	A2 picks up on this behavior and also responds with annoyance, 
	which we interpret as evidence that A1’s reaction 
	appeared strong not only from an outsider’s perspective 
	but also within the partner’s context.
\end{exmpl}

As already shown by the last two example, we only code an action as 
\Co{Hierarchical Behavior} if we see a reaction by the partner
that confirms that he interpreted the action as hierarchical. 
We will see either indications of disagreement with the behavior
or what we call \Co{Falling in Line}: submission to the partner's
assumed higher hierarchical position.
Both will be discussed below.

\begin{exmpl}[CA1 - Severe Reaction to a Mistake] \label{CA1_Belittling}
	This exchange happens early in the session. 
	The task is adding new functionality, including a corresponding GUI element. 
	C1 has started working on the task prior to the session. 
	He added a rudimentary element to the GUI that does not work properly.
	Before the present session, he commented out this GUI element.
	
	\DpartnerA{C1}{You can comment that in and then we'll see how it looks in the demo. 
	Then I can show you the effects that occur \DDialogCutOff} 
	
	\DpartnerBC{C2}{Yes, I can imagine that it won't work.
	  \DDescription{raises his voice} I can easily imagine it.}
	  \DCodeSub{Hierarchical Behavior}{Demeaning}
	
	\DpartnerAC{C1}{\DDescription{No reaction.}} \DCode{Falling in Line}
	
	\DpartnerB{C2}{\DDescription{C2 uncomments the line as suggested. 
	The IDE highlights the variable \texttt{ScaleRangePanel} in red.} 
	\texttt{ScaleRangePanel}\ldots}
	
	\DpartnerA{C1}{Why can't it \DDialogCutOff} 
	
	\DpartnerB{C2}{It must be called LabelScaleRangePanel. 
	\DDescription{Renames the variable. IDE highlighting disappears.} }
	
	\DpartnerA{C1}{Ah, I have renamed it. Yes, yes.}
	
	\DpartnerBC{C2}{Please wait a minute. 
	I want to check whether there's a comment in there or what it can do.}  
	\DCodeSub{Hierarchical Behavior}{Seizing Control}
	
	\DpartnerAC{C1}{\DDescription{No reaction.}} \DCode{Falling in Line}
	
	C2's first \emph{Hierarchical Behavior} utterance does not serve
    a productive purpose in the pair's dialogue.
	Rather, it suggests superiority. 
	
	C2's second \emph{Hierarchical Behavior} explicitly asks C1 to assume
	a subordinate role by not contributing actively. 
	In both cases, C1 does not react, which we interpret as 
	agreeing to a higher hierarchical position of C2.
\end{exmpl}
	
A different reaction of the partner,
challenging the implied pair relationship, is illustrated by
Example~\ref{AA1_Meta_Comm}:
	
\begin{exmpl}[AA1 - Lecturing \& Meta-Communication] \label{AA1_Meta_Comm}
  A1 and A2 discover source code they are unhappy with but that they have to
  deal with. 
  A2 is driving throughout the entire session AA1, 
  A1 performs \emph{Backseat-Driving} some of the time.
	
	\DpartnerA{A1}{Oh, there's a separate node for that, too. 
	And that one's even worse. 
	That’s where a special kind of request is cobbled together.} 
	\DDialogOm\footnote{We use this notation if dialogue has been omitted here to shorten the example.}
	
	\DpartnerB{A2}{Yeah, we can figure that out later.} 
	
	\DpartnerA{A1}{Yeah. Stick a \texttt{FIXME} on it for now \DDialogCutOff} 
	
	\DpartnerB{A2}{It has an object type. 
	But other than that, it doesn't know anything. 
	\DDescription{A2 scrolls through the source code ignoring A1's proposal.}}
	\DDialogOm
	
	\DpartnerA{A1}{Want to save that one for last?}
	
	\DpartnerB{A2}{Yeah.}
	
	\DpartnerA{A1}{Then just drop a \texttt{FIXME} here or something.} 

    \DpartnerB{A2}{\DDescription{Writes \texttt{FIXME} as a comment}}

    \DpartnerAC{A1}{So we don’t forget it. 
      We can also take care of that later in Eclipse.}
    
    \DCodeSub{Hierarchical Behavior}{Lecturing}

    \DpartnerBC{A2}{I know what \texttt{FIXME} is for.}
      \DCodeSub{Equalizing Behavior}{Meta-Communication}

    \DpartnerBC{A2}{
      \DDescription{Turns his head to A1 and laughs.} 
      Alright, next one?}
      
      \DCode{Equalizing Behavior}
	
	By ignoring A1's first proposal to annotate a \texttt{FIXME}, 
	A2 is thwarting A1's attempt to backseat-drive. 
	As A1 repeats his proposal and then explains it, 
	A2 reacts with meta-communication that rejects the
	\Co{Hierarchical Behavior}.
    Meta-communication of this type is rare.

    By laughing and immediately proposing to move on, 
    A2 also reduces tension and the pair continues to work together effectively.
    We call this \Co{Equalizing Behavior},
    which we will discuss in Section~\ref{sec:EqualBehav}.
\end{exmpl}

\subsection{Disengaging Behavior} \label{sec:DisBehav}

\HDef In \Co{Disengaging Behavior}, a partner gradually withdraws from
the session by not asking questions, not proposing anything,
and not challenging their partner's proposals.
\HRel \Co{Disengaging Behavior} is a common consequence of a \Co{Power Gap}.
By definition, it hampers \Co{Knowledge Transfer} which tends to result in
\Co{Knowledge Gaps} not getting smaller or even growing when the partner
makes progress.

\Co{Disengaging Behavior} consists mostly of non-actions and is therefore
hard to observe reliably.
For most cases, we use a helper concept, \Co{Unbalanced Process Agency},
which we will introduce in the next section.
The episode that ends in the above Example~\ref{DA2_PropSwitch}
is one of the few clear-cut cases of \Co{Disengaging Behavior}:
Over the course of several minutes, D3 makes fewer and fewer suggestions himself
and always agrees to those of his partner.
This withdrawal culminates in his refusal to become driver and he
pointed to his \Co{Knowledge Gap} as the justification.
We argue that in this case the \Co{Knowledge Gap} created a \Co{Power Gap}
(even without any \Co{Hierarchical Behavior} of the partner) which then
led to the \Co{Disengaging Behavior}.

Session DA2 from Example~\ref{DA2_PropSwitch} exhibits a vicious
cycle around \Co{Disengaging Behavior}
as can be seen in the following more complete version of the session:

\begin{exmpl}[DA2 - Disengaging Session Dynamic] \label{DA2_SessionD}
	DA2 starts with D3 driving and showing D4 around. 
	They are tasked with creating a new GUI element:
	a toolbar for a calendar view.
	D4 asks many project-related questions which D3 often cannot
	answer completely. He struggles with explaining certain details,
    attempts to cover up his ignorance, but D4 insists on precise
    understanding.
    That episode starts with D3 using an Eclipse Extension Point
    to view a GUI-Element.
	
	\DpartnerA{D4}{Are we still in the Eclipse view now? No?}
	
	\DpartnerBC{D3}{\DDescription{D3 shrugs and laughs.} 
		I have no idea about the extension points and all that. 
		I can’t give you any information on that.} 
		\DCode{knowledge gap marker} 
		
	[After encountering an error message when running the GUI, the pair
      switches to the Java code.]
	
	\DpartnerB{D3}{And they have this \texttt{setLicenceKey} here. 
		You’ll see it a few more times. 
		You basically have to put it in almost every component. 
		Each one wants its own licence key.} 
	
	\DpartnerA{D4}{What, you have to copy it each time or what?
      \DDescription{incredulous}}
	\DDialogOm
	
	\DpartnerB{D3}{Basically, yeah. Mhm. We can just search for it. 
	  \DDescription{D3 searches for \texttt{setLicenceKey}, but only finds one
        occurrence.} 
	  Here for example. Oh, that’s it already?
      \DDescription{D3 scrolls through the file.}
      Okay, normally there should be more components.} 
	
	\DpartnerA{D4}{So do we really have to set the licence key for every component?}
	
	\DpartnerBC{D3}{Uh\ldots mhm.
      \DDescription{D3 scrolls through the class again.} 
		That’s what I thought, but maybe that’s outdated. I’ll check.
		Okay, looks like it’s already changed.
		Alright, so it was just that one component after all.
		Would’ve surprised me, too, if every component needed it.
        \DDescription{laughs}} 
        
        \DCode{Hide Knowledge Gap} 
	
	A few minutes later D3 mentions his lacking knowledge regarding Java
    explicitly while explaining his work process to D4:
	
	\DpartnerBC{D3}{I
         took the demo that was closest to our own calendar,
         copied it in here, and 
		 then gradually adapted it piece-by-piece whenever issues came up. 
		 Because at the beginning, I had no idea about Java, and 
		 I had to start with something, right?} 
		 \DCode{knowledge gap marker} 
	
	\DpartnerA{D4}{Oh, so you hadn’t done any Java before that?}
	
	\DpartnerBC{D3}{No, I was completely inexperienced when it came to Java. 
		\DDescription{laughs} That’s why the code is total crap. 
		I mean, if you look here\ldots 
		I don’t even know how many lines we have by now, but\ldots}
		\DCode{knowledge gap marker} 
	
	\DpartnerA{D4}{God methods, god classes, right? \DDescription{laughs}}
	
	\DpartnerBC{D3}{Yes. 1917 lines of code. We have to refactor this some day.} 
	
	\DCode{knowledge gap marker} 
		
	[They discuss the solution approach where again D3 exhibits a low level
      of knowledge.
	D4 quickly understands the task and comes up with his own solution approach.
	D3 picks up on this.]
	
	\DpartnerBC{D3}{Have you already done anything with creating a toolbar like that?} 
	
	\DpartnerA{D4}{I actually read about it while I was on vacation. 
		\DDescription{Points to the book next to him.} 
		But if you don’t actually code along, you forget all of it. 
		I just reread it, though, at least the standard steps are in 
		there.
		But let’s first take a look at how they implemented [the previous toolbar] here.}
		
		\DCode{knowledge gap marker}\footnote{This is the only instance in the article where \CodeFormat{knowledge gap marker} refers not to the speaker's knowledge gap, but to that of the partner.}
		
	A few steps later, at 20 minutes into the session, D4 takes over the lead.
    We will later call this particular behavior \Co{Backseat-Driving}:
	
	\DpartnerA{D4}{Look at \texttt{plugin.xml}.} \DCodeSub{propose\_step}{low-level}
	
	\DpartnerBC{D3}{\DDescription{Opens \texttt{plugin.xml}}} 
	
	\DpartnerA{D4}{Under Extensions.} \DCodeSub{propose\_step}{low-level}
	
	\DpartnerBC{D3}{\DDescription{Navigates to the Extensions tab.}}
	\DDialogOm 
	
	\DpartnerA{D4}{\DDescription{Flips through is book.} 
      Is \texttt{ActionSet} included here?}
	
	\DpartnerBC{D3}{There are only standard Eclipse things in here.} 
	
	\DpartnerA{D4}{Show me the extensions.} \DCodeSub{propose\_step}{low-level}
	
	\DpartnerBC{D3}{\DDescription{Navigates to the Extensions tab again.}} 
	
	\DpartnerA{D4}{A bit further up.} \DCodeSub{propose\_step}{low-level}
	
	\DpartnerBC{D3}{\DDescription{Scrolls up.}} 
	
	\DpartnerA{D4}{Who did this? The to-do list?}
	
	
	
	
	[They bring in D5 and later D6 for help.
      D6 explains, D4 asks a lot of questions, and D3 sits quietly
      next to them.]
	
	After D6 left, we see the first distinct case of
    \Co{Disengaging Behavior} by D3. 
	
	\DpartnerBC{D3}{\DDescription{Laughs as D6 is leaving.} Okay. 
		\DDescription{Slides the keyboard over to D4} 
		I see you’ve kept up better than I have.} 
		
	\DCodeTwo{Disengaging Behavior}{knowledge gap marker}
	
	[D4 takes over and they start refactoring a class 
	by changing the name and altering the required objects for the class methods. 
	This creates nearly 100 error messages as the class 
	is used throughout the calendar view.] 
	
	D4 proceeds by creating a new class. 
	Their decision making is one-sided. 
	We will later refer to this as \Co{Solo-Driving}.
	Their exchange here is typical for the remainder of the session:	
	
	\DpartnerA{D4}{\DDescription{D4 opens the dialog to create a new class.} 
		Alright, I’ll do it differently this time.}
		\DCode{propose\_step}
	
	\DpartnerBC{D3}{\DDescription{Laughs.}} 
	
	\DpartnerA{D4}{I’ll just create a class like this, because, uh, 
		Eclipse only does crap \DDialogCutOff.
		\DDescription{D4 types \texttt{AbstractList} as the class name.}}
		
		\DCodeTwo{amend\_step}{propose\_design}
	
	\DpartnerBC{D3}{\texttt{AbstractListAction}.} \DCode{amend\_design}
	
	\DpartnerA{D4}{\DDescription{Adds \texttt{Action} to the class name and
	opens an Eclipse dialog to add an interface}}

	\DCode{amend\_design}
	
	\DpartnerBC{D3}{\texttt{ListAction}.} \DCode{amend\_design}
	
	\DpartnerA{D4}{
		\DDescription{Types \texttt{IView} in the search bar and selects the interface.}}
		
		\DCode{challange\_design}
	
	\DpartnerBC{D3}{Oh no, I\ldots} \DCode{agree\_design}
	
	\DpartnerA{D4}{\DDescription{Creates the class by confirming the dialog.}
	Soooooo, alright.}
	\DDialogOm
	
	\DpartnerA{D4}{\DDescription{Deletes the \texttt{getName} method.} 
		We don't actually need that. 
		\DDescription{D4 continues by jumping into the empty execute method.}
		If\ldots \DDescription{Types an if-statement.} Abstract\ldots}
		\DCode{propose\_design}
	
	\DpartnerBC{D3}{\texttt{AbstractListPreview}.} \DCode{amend\_design}
	
	\DpartnerA{D4}{
		\DDescription{Selects \texttt{AbstractListPreviewContainer} 
			and start writing a new method.}
		What should we call the abstract method that’s being called?
		\DDescription{Starts typing \texttt{execute}.}}
		\DCode{ask\_design}
	
	\DpartnerBC{D3}{Mhm. \texttt{ExecuteALAPC}. 
		\DDescription{Laughs.}} \DCode{propose\_design}
	
	\DpartnerA{D4}{\DDescription{Laughs.} Uhm.}
	
	\DpartnerBC{D3}{Execute Abstract List and Preview.} \DCode{amend\_design}
	
	\DpartnerA{D4}{\DDescription{Silence.}}
	
	\DpartnerBC{D3}{Even if that’s a bit long, but\ldots} \DCode{amend\_design}
	
	\DpartnerA{D4}{\DDescription{D4 ignores D3 and enters the name 
			\texttt{internalExecute} instead.}}
	
	\DCode{challange\_design}
	
	[D4 continues adding logic to the class. D3 stays mostly reticent. 
	He occasionally helps mostly by naming the required class for an object. 
	D4 rarely asks questions but looks directly in the code for an answer. 
	D3 only asked once for a technical explanation for D4's doing.]
	
	After finishing the new class D4 spends the next 8 minutes fixing various error messages that resulted from their previous refactoring step.
	D4 institutes this by saying:
	
	\DpartnerA{D4}{Alright, then I'll start with the dirty work.}
	
	\DpartnerBC{D3}{\DDescription{Laughs.}} 
	
	[D4 continues with the refactoring. 
	D3 is mostly passive and watches D4. 
	He occasional helps by providing the class names.]
	
	After fixing the first few error messages, D4 proposes a 
	Driver/Observer switch. 
	Please refer to Example~\ref{DA2_PropSwitch} to see their exchange. 
	
%
%
%
 
    [D4 continues with fixing the error messages. D3 proactively asks one question 
    regarding abstract classes in Java.]
    
    Approx. 10 minutes later D4 employs Meta-Communication:
    
    \DpartnerA{D4}{So just tell me if you want to take over again. 
    	\DDescription{Looks at D3 and points to the keyboard.}}
    	\DCode{Equalizing Behavior}
    
    \DpartnerBC{D3}{No, you keep going for now. 
    	Once I’m fully back into it, I’ll shout.} 
    	
    [Fixing the error messages becomes a repetitive task.]
    
    Approx. 5 minutes later D4 again proposes a Driver/Observer-Switch
    after explicitly mentioning the repetitiveness of the remaining tasks:
    
    \DpartnerA{D4}{So, this might get a bit boring now.}    
    \DDialogOm
    
    \DpartnerA{D4}{Yeah, but if you want, you can also\ldots}
    
    \DpartnerBC{D3}{If you stay for another minute. \DDescription{They change seats and Driver/Observer roles}}    
    \DDialogOm 
    
    \DpartnerA{D4}{Nah, I don’t know, let's do ten classes at a time. 
    	You’ll just make ten more classes, than I'll do ten.}
    
    [They continue with D3 as Driver with D4 guiding them through the tasks.
    D3 asks more questions and their process becomes less one-sided. 
    They soon face more challenges than expected 
    which leads D4 to taking back the driver role after 10 minutes. 
    After resolving these challenges they switch roles 
    for the remainder of the session
    until all error messages are resolved.]  
    
    Here we described the whole session DA2. 
    In the middle of the episode we see \Co{Disengaging Behavior} 
    by D3 which is highlighted by \Co{One-sided Decisions} and in 
    turn \emph{Unbalanced Process Agency}, which we will both discuss 
    in Section~\ref{sec:ProcessAgency}. 
    This \Co{Disengaging Behavior} is probably driven by D3's intention 
    to hide or overplay his \Co{Knowledge Gaps}. 
    We see multiple markers for this throughout the session and some
    active attempts of overplaying or hiding them.  
    As already discussed, this also becomes visible when D3 refuses to drive.   
    This behavior increases the \Co{Knowledge Gap} as D3 
    presumably holds back asking questions while D4 increases his 
    understanding either by looking through the code or 
    discussing the problem with D5 and D6.     
    D3 becomes more active as the task becomes simpler 
    (which D4 directly mentions) and D3 takes over the Driver role. 
    Through asking more questions and actively engaging we believe 
    that D3 decreased his \emph{Knowledge Gap} and the session becomes
    less dysfunctional.    
\end{exmpl} 


We see how the \Co{Disengaging Behavior}
makes \Co{Knowledge Transfer} almost disappear,
which (as work goes on) results in an increasing \Co{Knowledge Gap},
which widens the \Co{Power Gap}.
In the end, the \Co{Power Gap} is so large that D3 even refuses
something as ordinary as a driver change.

We are sometimes able to understand what should have happened instead
of a \Co{Disengaging Behavior}.
In most such cases, pair members should have asked a question for reducing
their \Co{Knowledge Gap}.
We call this \Co{Not Pulling Important Knowledge}.
It occurs for instance in Example~\ref{PA3_AssertSame} when
P3 sticks to \mbox{assertEquals} because he does not know how to use
\mbox{assertSame}, for which P1 made a clear case.
A second case happens a while later in the same session and corroborates
the diagnosis.

We believe that the motivation behind \Co{Not Pulling Important Knowledge}
is \Co{Hiding Knowledge Gaps}.
Corresponding behavior was described in two of the interviews we held:
\Qby{I5}{QRzDKTAnnoy2}{That's why I often ask people,
  \say{Do you know how this and that works, do you know the technology,
    this working method or anything else?}
  And funnily enough, it's the supposed seniors who can't admit that
  they don't know this.
  They then start to get stressed and feel tested.
}\\
\Qby{I1}{QteDistance1}{[My partner reacted annoyed to suggestions, which] was
  also a signal for me [\ldots] to perhaps make up for a lack of understanding
  before I asked new questions[\ldots].
  Yes, that always created a bit of distance [between us] [\ldots].}

The last sentence states an increasing \Co{Power Gap},
which is growing, at least in part, because of
the interviewee's \Co{Disengaging Behavior}
(\say{before I asked new questions}),
which happened presumably because of the partner's negative emotions
(\say{reacted annoyed}),
which resulted in a motivation to hide knowledge gaps
(\say{a signal for me\ldots}).

A behavior somewhere between \Co{Defensive Behavior}
(in the form of \Co{Minimizing Communication}) and
\Co{Disengaging Behavior}
(in the form of \Co{Hiding Knowledge Gaps}) is a pattern of
\Co{Asking Indirectly}. Examples:
\Qby{P3}{QresponseContent}{I don't know exactly how responseContent works here.} or
\Qby{P3}{Qaskindirect2}{But I can't remember how to call the service
  in Symfony at the moment.}  

\subsection{Equalizing Behavior} \label{sec:EqualBehav}

\HDef \Co{Equalizing Behavior} is any utterance (or action) that
makes an existing \Co{Power Gap} smaller (\Co{Reductive}) or
avoids the creation of a \Co{Power Gap} (\Co{Preventive}).
\Co{Equalizing Behavior} is available
to both partners, the one with the higher hierarchical position (\Co{from-above})
and the inferior one (\Co{from-below}).
\HRel \Co{Equalizing Behavior} has the opposite effect from
\Co{Hierarchical Behavior}.

Except for the possible \Co{from-above}/\Co{from-below} asymmetry,
the operationalization of this concept works along the same lines as for
\Co{Hierarchical Behavior}.

For a first example, see again Example~\ref{AA1_Meta_Comm}
where A2 employs \Co{Meta-Communication} to counter A1's
\Co{Hierarchical Behavior}. 
This is \Co{Reductive Equalizing Behavior}. 
He immediately proceeds with reducing tension.
This particular tension-reducing behavior is at once
\Co{from-below} (because of the partner's previous \Co{Hierarchical Behavior})
and \Co{from-above}.
It is \Co{from-above} because this pair is jostling constantly throughout the
first half of the session:
The two have different work styles and both exhibit
\Co{Hierarchical Behavior} frequently.
In this session, two \Co{Power Gaps} exist at once,
fluctuating in their acute amount of relevance.
Both partners are the superior person for one of the \Co{Power Gaps}
and the inferior person for the other.

Example~\ref{CA5_Agree_to_Disagree} shows a similar, but much more successful
pair.
They, too, have different work styles and therefore lots of reasons
for tension.
But they know each other well and perform frequent
\Co{Preventive Equalizing Behavior} so skillfully
that no \Co{Power Gap} ever becomes visible:

\begin{exmpl}[CA5 - Agree to Disagree] \label{CA5_Agree_to_Disagree}
	C3, who is currently editing the code, makes a proposal on using ensures.
    C4 prefers a style that relies on tests, not on ensures.
	
	\DpartnerB{C3}{You know what I'm going to write now. 
		\DDescription{Writes \texttt{ensure}.}}
      
      \DCode{Preventive Equalizing Behavior}
	
	\DpartnerAC{C4}{No no, please please no. 
		\DDescription{Smiles.} 
		Let's write tests. 
		\DDescription{Looks to C3.} 
		Then you don't want those Ensures there. 
		\DDescription{Laughs.}} 
      
      \DCode{Preventive Equalizing Behavior}
	
	\DpartnerBC{C3}{Will they hurt you if I write them down now?} 
      
      \DCode{Preventive Equalizing Behavior}
	
	\DpartnerAC{C4}{\DDescription{Laughing.} 
		Yeah, totally. 
		But it's fine if it doesn't hurt you when I remove them when
        I'm testing.}
      \DCode{Preventive Equalizing Behavior}
	
	\DpartnerB{C3}{When you're testing, if they get in the way of your testing. 
		But as long as they're not in the way?} 
	
	\DpartnerAC{C4}{\DDescription{Smiles.} 
		They are in the way.}
      \DCode{Preventive Equalizing Behavior}
	
	\DpartnerBC{C3}{\DDescription{Uses \texttt{ensure} four times. 
			And looks to C4, smiling.}} 
      
      \DCode{Preventive Equalizing Behavior}
	
	\DpartnerAC{C4}{We'll never agree on that. 
		\DDescription{Smiles.} 
		Never!} 
      
      \DCode{Preventive Equalizing Behavior}
	
	\noindent [20 minutes later; C4 is driving now. They revisit the topic.]
	
	\DpartnerB{C3}{Hey?}
	
	\DpartnerA{C4}{Mhm?} 
	
	\DpartnerB{C3}{Will you gift me an \texttt{ensure}?} 
      \DCode{Preventive Equalizing Behavior}
	
	\DpartnerAC{C4}{Oh my God, you are a sadist! 
		\DDescription{Writes \texttt{ensure}.} 
		Shall we talk it over [over a beer] sometime soon?} 
      \DCode{Preventive Equalizing Behavior}
	
	\DDialogOm 
	
	\DpartnerB{C3}{If you like, sure we can do that.}
      \DCode{Preventive Equalizing Behavior}
	
	Although C3 and C4 disagree fundamentally on the use of \texttt{ensure}, 
	they communicate in a way that deescalates any tension that might result
	by laughing about their disagreement and by explicitly asking for approval,
	no matter who of them is driving. 
	The partner also always reacts in a deescalating way,
    which is evidence that they registered and acknowledged the 
    \Co{Preventive Equalizing Behavior} intention of their partner.
\end{exmpl}

One of our interviewees  (in continuation
of quote~\ref{QRzDKTAnnoy2}) described
\Co{from-above Reductive Equalizing Behavior}.
\Qby{I5}{QpEqual2}{making it completely
    okay to ask questions [\ldots]. If you notice [Disengaging
    Behavior], you can definitely say something like, \say{I get the
    impression that you're feeling a bit stressed.} [...] I’m not here
    to grade you or anything.}

\subsection{Unbalanced Process Agency} \label{sec:ProcessAgency}

Those \Co{Power Gaps} that result in \Co{Disengaging Behavior} are difficult
to diagnose because \Co{Disengaging Behavior} itself is difficult to observe.
We found, however, that \Co{Unbalanced Process Agency} always strongly suggests
the presence of a \Co{Power Gap} and, if no other symptoms are visible,
the ``presence'' of \Co{Disengaging Behavior}.

We explain \Co{Unbalanced Process Agency} by means of two auxiliary concepts,
\Co{Shared Decision} and \Co{One-sided Decision}.
\Co{Unbalanced Process Agency} itself is not an auxiliary concept, however,
because it is useful for explaining pair programming skill.

Pair programming involves constant decision making.
In a good pair session, both partners meaningfully influence almost
every decision because this helps to keep \Co{Togetherness} high.
We call such decisions \Co{Shared Decisions}.

In a problematic session, many decisions are made by one partner alone
(we will explain the difference in detail below), which we call
\Co{One-sided Decision}.
If many decisions in a row are made one-sidedly by the same partner,
we diagnose \Co{Unbalanced Process Agency}.

A developer can influence a decision by proposing something or by
disagreeing with, challenging, or amending a proposal.
Or influence it by adding knowledge in the form of hypotheses and findings.
We conceptualize such elementary behaviors using Salinger \& Prechelt's
Base Layer for Pair Programming~\cite{SalPre13}. 

\Co{One-sided Decisions} occur in two different forms, 
depending on whether the partner controlling the decision-making is 
controlling computer input (``driver'') or not.
If the main decision-maker is the driver, we call the behavior
\Co{Solo-Driving} (otherwise \Co{Backseat-Driving}):

\begin{exmpl}[DA4 - Solo-Driving] \label{DA4_Solo-Driving}
  D5 and D6 are fixing a bug in a compare function of a class C
  that contains a list of other objects.
  Currently, two different C objects that hold the same list are
  considered equal, which is unintended behavior.
  This example starts with D6 proposing to convert the lists to arrays
  for avoiding this behavior.
	
  \DpartnerA{D6}{But if we have an array, if we use an array,
    then the list’s equals won’t be used. Let’s try that.}
	
	\DpartnerB{D5}{Yeah, fine. Let’s do it.} 
	\DDialogOm
	
	\DpartnerAC{D6}{\DDescription{Starts to edit the code by changing
        the class of the objects from list to array.}
      I’m just going to shamelessly write all of this down, okay?}
    
    \DCode{propose\_strategy}
	
	\DpartnerBC{D5}{Mhm.} \DCode{agree\_strategy}
	
	\DpartnerAC{D6}{To\ldots \DDescription{Converts an existing list object
        to array by writing \texttt{toArray()}.}
      I’ll get the \texttt{ArrayList} here.} \DCode{propose\_design}
	
	\DpartnerBC{D5}{Mhm.} \DCode{agree\_design}
	
	\DpartnerAC{D6}{Apply \texttt{toArray}. And convert this variable here to\ldots\
      \DDescription{adds \texttt{toArray}.}} 
      
      \DCode{propose\_design}
	
	\noindent [D6 continues to verbalize his thinking and actions while he
      further alters the methods and runs the application to test it.]
	
	D6 makes all decisions alone, sometimes even editing the code before
    verbalizing the proposal to do it. 
    D5 is mostly passive, only signaling agreement (with a minimum of speech).
    This constitutes \Co{Solo-Driving} (which D6 had even proposed explicitly
    by saying
	\say{I’m just going to shamelessly write all of this down, okay?}).
\end{exmpl}

\Co{One-sided Decisions} can be made by the observer as well.
If the driver only implements the observer's proposals and those proposals are
highly detailed and specific (\Co{low-level}), we call the behavior
\Co{Backseat-Driving}\footnote{The idea of Backseat-Driving was first mentioned by Jones and Fleming~\cite{JonFle13}. To the best of our knowledge, we are the first to formalize it.}:

\begin{exmpl}[DA5 - Backseat-Driving] \label{DA5_Backseat}
  D2 and D8 just started to expand an existing test case.
  D8 is driving.
	
  \DpartnerBC{D8}{For\ldots\ \DDescription{Writes a for-loop.}
    So, I'm going through it once, and if I find one -- what do I do with it?}
  \DCodeTwo{propose\_design}{ask\_design}
	
	\DpartnerAC{D2}{It should already have found exactly one.} \DCode{disagree\_design}
	
	\DpartnerB{D8}{Oh, right. \DDescription{Deletes for-loop.}} 
	\DDialogOm
	
	\DpartnerBC{D8}{Then we say \texttt{assert}?}
    \DCodeTwo{propose\_design}{follow-up\_inference}
	
	\DpartnerAC{D2}{Exactly.} 	
	\DDialogOm
	\DCode{agree\_design}
	
	\DpartnerB{D8}{\DDescription{Writes \texttt{assert}. 
        The IDE shows different Assert functions.}} 
	
	\DpartnerAC{D2}{\texttt{assertEquals}.} \DCodeSub{propose\_design}{low-level}
	
	\DpartnerBC{D8}{\DDescription{Selects \texttt{assertEquals}.
        This function requires a \texttt{boolean expected} and a
        \texttt{boolean actual}.} \texttt{Boolean expected}?} \DCode{ask\_design}
	
	\DpartnerAC{D2}{\texttt{True}.} \DCodeSub{propose\_design}{low-level}
	
	\DpartnerBC{D8}{And is it supposed to be a Boolean here, too? No?}
    \DCode{ask\_design}
	
	\DpartnerAC{D2}{\texttt{activities.size equals 1}.} 
	\DDialogOm 
	\DCodeSub{propose\_design}{low-level}	
	
	\DpartnerB{D8}{Okay, we’ve got that now. Let’s move on.}
    
    \DCodeTwo{explain\_completion}{propose\_strategy}
	
	\DpartnerA{D2}{We can test it first if you want.}
    \DCode{challenge\_strategy}
	
	\noindent [D2 continues to guide D8 through the IDE in detail:
      set a breakpoint; run the test case.]
	
	Although D8 proposes to use Assert here, we don't count it as
    a proposal that influences the decision-making, as it is
    the obvious next step in the current test code context
    (\Co{follow-up\_inference}).
    As all other proposals that influence the episode (and are not
    immediately discarded) stem from D2 and are \Co{low-level},
    the whole decision-making episode is \Co{Backseat-Driving}.
\end{exmpl}

We defined a \Co{Shared Decision} by both developers contributing
in a meaningful way. 
But what does \say{meaningful} mean? 
We take together all utterances and actions of one partner during a
decision-making episode and ask:
\say{Would the other developer be able to reach a similar conclusion
  in a similar time without these utterances?}
If the answer is no, we call it a \Co{Shared Decision}. 
If the answer is yes, but both partners still had some influence, 
we cannot decide between a \Co{Shared} vs a \Co{One-sided Decision}.
As most decisions in pair programming are small (often only a proposal
followed by an agreement), 
\Co{One-sided Decisions} are common and usually unproblematic. 

\HDef If and only if multiple decisions in short succession are
\Co{One-sided Decisions} and are made by the same developer,
we characterize the whole stretch as \Co{Unbalanced Process Agency}. 
\Co{Unbalanced Process Agency} can last for only a minute or
for a large part of a session.

See again Example~\ref{DA2_SessionD}. 

\subsection{A Grounded Theory of Power Gaps in PP} \label{sec:GT}

The above sections already mention most parts of the 
\Co{Grounded Theory of Power Gaps in Pair Programming} in the form of
concept definitions and relationships between these.
Please study the visualization of our grounded theory in 
Figure~\ref{fig_GT} and reactivate your understanding of all those
concepts and relationships already discussed above.
The present section will add further concepts and relationships,
refer to the examples used for grounding,
and extend the grounding (via published
literature and argumentation) in those cases where it is missing in the
examples and discussion in previous sections.

\usetikzlibrary {positioning}
\usetikzlibrary{shapes.geometric, arrows, backgrounds, fit, calc}
\tikzstyle{SuccessFactor} = [circle, minimum width=1.2cm, text width=1.1cm, text centered, inner sep=1pt, draw=black, fill=green!30, font=\scriptsize]
\tikzstyle{RiskFactor} = [circle, minimum width=1.2 cm, text width=1.1cm, text centered, inner sep=1pt, draw=black, fill=red!30, font=\scriptsize]

\tikzstyle{Outcome} = [circle, minimum width=1.2cm, text width=1.1cm, text centered, inner sep=1pt, draw=black, fill=blue!15, font=\scriptsize, thick]

\tikzstyle{Behavior} = [ellipse, minimum width=1.5cm, minimum height=0.8cm, text width=1.3cm, inner sep=1pt, text centered, draw=black, fill=gray!10, font=\scriptsize]
\tikzstyle{Legend} = [rectangle, minimum width=2cm, minimum height=1.5cm, text width=2cm, inner sep=4pt, text centered, draw=black, font=\scriptsize]
\tikzstyle{posArrowL} = [green!50!black!50, , thick, line width=1mm,->,>=stealth]
\tikzstyle{posArrowS} = [green!50!black!50, ->,line width=0.3mm,>=stealth]
\tikzstyle{negArrowL} = [red!50!black!60, thick, line width=1mm,->,>=stealth]
\tikzstyle{negArrowS} = [red!50!black!60,->,line width=0.3mm,>=stealth]
\tikzstyle{posArrowLB} = [green!50!black!50, , thick, line width=1mm,<->,>=stealth]

\tikzstyle{subConceptA} = [o-stealth, black, thick, dashed]

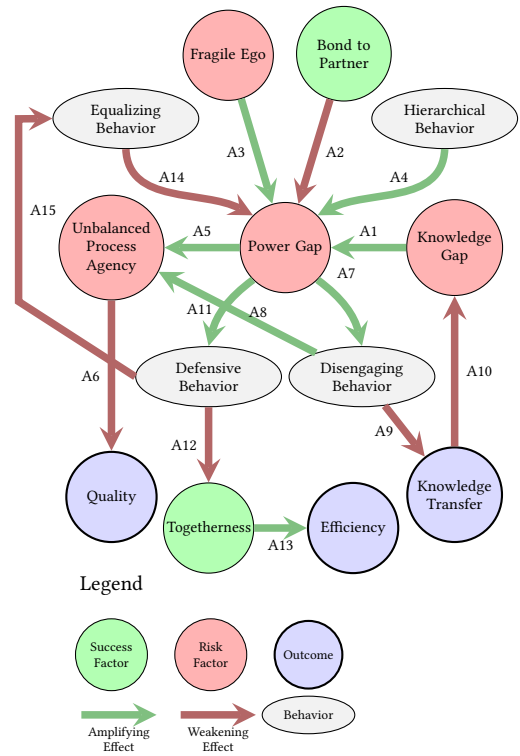
\begin{figure}[tbh]
	\begin{tikzpicture}
		\node (PGap) at (0,0) [RiskFactor] {Power Gap};
		
		\node (KGap) [RiskFactor, right = of PGap] {Knowledge Gap};
		
		

		\node (HBehav) [Behavior, above right = of PGap] {Hierarchical Behavior};
		
		\node (EBehav) [Behavior, above left = of PGap] {Equalizing Behavior};
		
		\node (DBehav) [Behavior, below left = 1 and -0.1 of PGap] {Defensive Behavior};
		
		\node (DisBehav) [Behavior, below right = 1 and -0.1 of PGap] {Disengaging Behavior};
		
		\node (KT) [Outcome, below = 2 of KGap] {Knowledge Transfer};

		\node (UPA) [RiskFactor, left = of PGap] {Unbalanced Process Agency};
		
		\node (Tog) [SuccessFactor, below = 1 of DBehav] {Togetherness};
		
		\node (Quality) [Outcome, below = 2 of UPA] {Quality};
		
		\node (Efficiency) [Outcome, right = 0.7 of Tog] {Efficiency};
		
		
		\draw [posArrowL] (KGap) to node[black, font=\scriptsize, midway, above] {\refA{A-KGap-PGap}} (PGap);
		
		
		
		
		\draw [posArrowL, bend left=60, in=180] (HBehav.south) to node[black, font=\scriptsize, midway, above] {\refA{A-HBehav-PGap}} (PGap.north east);
		
		\draw [negArrowL, bend right=60, in=180] (EBehav.south) to node[black, font=\scriptsize, midway, above] {\refA{A-EBehav-PGap}} (PGap.north west);
		
		
		\draw [negArrowL] (DisBehav) to node[black, font=\scriptsize, midway, left] {\refA{A-DisBehav-KT}} (KT);
		
		\draw [posArrowL] (DisBehav) to node[black, font=\scriptsize, midway, right] {\refA{A-DisBehav-UPA}} (UPA);
		
		
		\draw [posArrowL, bend right =20] (PGap.south west) to node[black, font=\scriptsize, midway, left] {\refA{A-PGap-DefBehav}} (DBehav.north);
		
		\draw [posArrowL, bend left=20] (PGap.south east) to node[black, font=\scriptsize, midway, above = 0.25] {\refA{A-PGap-DisBehav}} (DisBehav.north);
		
		\draw [posArrowL] (PGap) to node[black, font=\scriptsize, midway, above] {\refA{A-PGap-UPA}} (UPA);
		
		\draw [negArrowL] (UPA) to node[black, font=\scriptsize, midway, left] {\refA{A-UPA-Qual}} (Quality);
		
		
		\draw [negArrowL] (DBehav) to node[black, font=\scriptsize, midway, left] {\refA{A-DefBehav-Tog}} (Tog);
		
		
		\draw [posArrowL] (Tog) to node[black, font=\scriptsize, midway, below] {\refA{A-Tog-Ef}} (Efficiency);
		
		\draw [negArrowL] (KT) to node[black, font=\scriptsize, midway, right] {\refA{A-KT-KGap}} (KGap);
		
		
		\draw [negArrowL] (DBehav.west) --++ (-1.55,1) --++ (0,2.43) node[black, font=\scriptsize, midway, right] {\refA{A-DefBehav-EBehav}} to  (EBehav.west);
		
		

		\node (LegSF) [SuccessFactor, minimum width=9mm, text width=8mm, font=\tiny, below = 1 of Quality] {Success Factor};
		\node (LegRF) [RiskFactor, minimum width=9mm, text width=8mm, font=\tiny, right = 0.35 of LegSF] {Risk Factor};
		\node (LegOut) [Outcome, minimum width=9mm, text width=8mm, font=\tiny, right = 0.35 of LegRF] {Outcome};
		\node (LegBe) [Behavior, minimum width=10mm, minimum height=5mm, text width=8mm, font=\tiny, below = 0.08 of LegOut, yshift = 0 mm] {Behavior};
		\draw [posArrowL, font=\tiny] ($ (LegSF) + (-0.4,-0.8) $) to node (BGA)[black, anchor=north, xshift=0mm, align = center] {Amplifying \\ Effect} ($ (LegSF) + (0.6,-0.8) $);
		\draw [negArrowL, font=\tiny] ($ (LegRF) + (-0.4,-0.8) $) to node (BRA)[black, anchor=north, xshift= 0 mm, align = center] {Weakening \\ Effect} ($ (LegRF) + (0.6,-0.8) $);
		\node (LegLabel) [above = 0.2 of LegSF, font=\small] {Legend};
		
	\end{tikzpicture}
	\caption{Grounded Theory on the Power Gap in Pair Programming. See the description in Sections~\ref{sec:GT}}
	\label{fig_GT}
\end{figure}

The context is a professional pair programming session with two developers. 
Almost always, one developer has more task-relevant knowledge than the other,
i.e., there is a \Co{Knowledge Gap}.

This \Co{Knowledge Gap} tends to lead to a \Co{Power Gap}
\newA{A-KGap-PGap}{sec:KGap}{
  examples~\ref{DA2_PropSwitch}, \ref{PA3_AssertSame}, \ref{DA2_SessionD},
  and the sessions of examples
  \ref{CA1_Belittling}, \ref{AA1_Meta_Comm}, \ref{DA5_Backseat}}. 
%
%
\Co{Hierarchical Behavior} increases the \Co{Power Gap} 
\newA{A-HBehav-PGap}{sec:HBehav}{
  examples~\ref{PA3_AssertSame}, \ref{AA1_MiniObject}, \ref{CA1_Belittling}, \ref{AA1_Meta_Comm}}.
A big \Co{Power Gap} can lead to \Co{Unbalanced Process Agency} 
where mostly one partner is doing the decision-making
\newA{A-PGap-UPA}{sec:ProcessAgency}{
  examples~\ref{DA2_SessionD},
  and the sessions of examples~\ref{PA3_AssertSame}, \ref{CA1_Belittling}, \ref{DA4_Solo-Driving}},
which in turn can hurt the \Co{Quality} of the task outcome
\newA{A-UPA-Qual}{sec:ProcessAgency}{
  research that pair programming in general has a positive effect on quality
  \cite{HanDybAri09, FuGraBro17, VanLas05, Zac11, AriGalDyb07, VanMan13}}.

A big \Co{Power Gap} can lead the \Co{subordinate} partner into
\Co{Disengaging Behavior}
\newA{A-PGap-DisBehav}{sec:DisBehav}{
  examples~\ref{DA2_PropSwitch}, \ref{DA2_SessionD},
  and the session of example \ref{DA4_Solo-Driving}},
which creates or increases \Co{Unbalanced Process Agency}
\newA{A-DisBehav-UPA}{sec:DisBehav}{
  examples~\ref{DA2_PropSwitch}, \ref{DA2_SessionD}, \ref{DA4_Solo-Driving}}
and hurts \Co{Knowledge Transfer}
\newA{A-DisBehav-KT}{sec:DisBehav}{
  definition of \Co{Disengaging Behavior}},
which increases or keeps up the \Co{Knowledge Gap}
\newA{A-KT-KGap}{sec:DisBehav}{
  Examples~\ref{DA2_SessionD}},
which tends to keep up the \Co{Power Gap}
(with the grounding seen above).

A big \Co{Power Gap} can also lead the \Co{subordinate} partner into
\Co{Defensive Behavior}
\newA{A-PGap-DefBehav}{sec:DefBehav}{
  Example~\ref{PA3_AssertSame}},
which reduces \Co{Togetherness}
\newA{A-DefBehav-Tog}{sec:DefBehav}{
  definition of \Co{Defensive Behavior}, in particular
  \Co{Minimizing Communication}},
which can lead to breakdowns, 
which hurt the \Co{Efficiency} of the pair programming process
\newA{A-Tog-Ef}{sec:DefBehav}{
  research that links \Co{Togetherness} with Process Fluency \cite{Zie20}}.

\Co{Equalizing Behavior} works as an antidote by reducing the \Co{Power Gap}
\newA{A-EBehav-PGap}{sec:EqualBehav}{
  examples~\ref{AA1_Meta_Comm}, \ref{DA2_SessionD},
  \ref{CA5_Agree_to_Disagree}}.
\Co{Defensive Behavior} makes \Co{Equalizing Behavior} less likely
\newA{A-DefBehav-EBehav}{sec:DefBehav}{
  the session of Example~\ref{PA3_AssertSame}}.

\subsection{Managing Power Gaps as an Element of Pair Programming Skill} \label{sec:ppskill}

Each instance of \Co{Equalizing Behavior}
(whether \Co{from-above} or \Co{from-below}) is an instance of
applied pair programming skill.
Each case of \Co{Hierarchical Behavior} (\Co{from-above})
and likewise each case of
\Co{Defensive Behavior} or \Co{Disengaging Behavior} (\Co{from-below})
are instances of insufficient pair programming skill.

Do not be misled: Our article needed to explain the concepts and so contains
much more material about non-skillful behavior than about skillful behavior.
But coping with \Co{Knowledge Gaps} is the bread and butter of pair programmers
and we see lots of pairs with large \Co{Knowledge Gaps} where little or no
dysfunctional behavior occurs.

How to teach and learn this part of pair programming skill is a topic
for future work.

\section{Limitations} \label{sec_lim}

Grounding: Several of our main concepts are not directly observable
because they refer to subjective perceptions.
The upside of this decision is the high relevance of our Grounded Theory:
It explains phenomena involving emotions, a topic area that many software
developers avoid thinking about which was also mentioned by one of our interviewees: 
\Qby{I6}{QABMeta}{[E]verything related to soft skills tends to be pushed more by engineering managers
  rather than by other developers. [\ldots]
  Between developers themselves, [meta-discussion] happens rather rarely.}.

But can these concepts be grounded solidly?
Yes. Much of our text in Section~\ref{sec:res} is used for explaining
how we made sure the diagnosis of these concepts is tangible and reliable
despite the subjectivity they describe.
The resulting grounding is documented in detail by the many cross references
we provide, so the reader can trace every instance they want to question.

Theoretical Sampling: In interview-based Grounded Theory works,
Theoretical Sampling means to lead additional interviews and ask those
questions that the analysis-so-far has raised.
For the kinds of phenomena investigated here and observational data,
this is impossible: One cannot know what will happen in the next pair session,
no matter how one would select the pair and the setting -- which is impractical
to begin with.
The only solution is to have a large collection of sessions in store,
know the rough shape of each, and select from that collection in order to
satisfy one's Theoretical Sampling needs approximately.
In our case, the required material is at the episode level and the repository
contains over 2000 episodes.
In this set, we always found suitable material when we needed it.

Theoretical Saturation: Due to the Theoretical Sampling, our Grounded 
Theory covers all phenomena seen in the repository of pair 
programming sessions we used.
More phenomena likely exist in the overall universe of pair programming.
However, we do not expect that any of these would require an extension
(let alone a modification) of the Grounded Theory presented above:
The degree of generality of each of its concepts is so high that they
will cover such additional behaviors as long as the aspect of
interest is \Co{Power Gap} dynamic.

Representativeness:
The question how typical the phenomena described herein are for pairs
of different types is irrelevant for the validity of
our research.
Our research is an existence proof.
It does not become invalid by additional things existing as well.

The recorded pair programming sessions we analyzed span over more than a 
decade and the pairs employ different technology accordingly.
The pair programming behaviors analyzed here, however, do not change over time:
For none of the phenomena described did it ever make a difference
whether the session we were looking at was old or young;
our results appear to describe timeless phenomena.
Also, there is currently no evidence that solo programming with an 
LMM-based AI-tool will replace pair programming in all 
cases~\cite{Imai22, LyuWanSun25, SarGorNeg22, SimParTam24, FanLiuZha25} 
which indicates that pair 
programming (and research for understanding it) will continue to be relevant.

\section{Related Work}\label{SecRelatedWork}

The first discussion of pair programming skill focused on
knowledge transfer skill only~\cite{ZiePre14},
the second was concerned with maintaining \Co{Togetherness} and
\Co{Fluency}~\cite{ZiePre21}.

Plonka et al. studied disengagement in pair programming by analyzing
21 industrial sessions~\cite{PloShaLin12}. 
They identified social pressure as one of five reasons for disengagement
where novice developers who had \say{less knowledge to solve the task
  at hand [...] were disengaged during the PP sessions;
  they avoided driving and they stopped asking and replying to questions.} 
Through interviews with the developers they found that
\say{less experienced developers can be worried about slowing
  their partners down or that they may look stupid.}
Besides acceptable disengagement for example where the pair was
\say{solving a simple task that does not require input from both developers},
Plonka et al. also identified developers dropping out of the session as
a result of social pressure and time pressure. 
They found it harmful, as the developers were
\say{not able to follow their partner’s explanations or activities}
which \say{undermines knowledge transfer}.
Our \Co{Disengaging Behavior} concept captures many of the same events.
The connection to \Co{Power Gap} provides a richer explanation than
previously available and \Co{Equalizing Behavior} shows a way out.

Zieris and Prechelt find \say{there is usually no pair member who is
  more knowledgeable in all relevant areas} \cite{ZiePre16}.
Insofar as this is correct, it suggests the near-constant danger of
two-sided \Co{Power Gaps}, but also chances for
two-sided \Co{Equalizing Behavior}.

Chong and Hurlbutt conducted an ethnographic observation of pair programming
in two teams over the course of multiple weeks \cite{ChoHur07}.  
They describe what we call a (probable) \Co{Power Gap} with
(certain) \Co{Unbalanced Process Agency}:
\say{The gaps in expertise between the programmers on Team B clearly influenced
  pair programming interactions on the team.
  On Team B, the member of the pair with greater expertise drove
  the bulk of the programming discussions}.

Begel \& Nagappan surveyed 487 developers on pair programming at
Microsoft~\cite{BegNag08}. 
They found that a good pair should have \say{compatible personalities}, and
\say{no ego}. 
Our work does not attempt to operationalize personality,
but \say{no ego} presumably means a very low incidence of
\Co{Hierarchical Behavior},
\Co{Defensive Behavior}, and
\Co{Disengaging Behavior}
and a good chance of \Co{Equalizing Behavior} whenever a \Co{Power Gap}
threatens to appear nevertheless.

Hannay et al. conduct a meta-analysis of controlled experiments on the
effects of pair programming vs solo programming~\cite{HanDybAri09}
and find high variance between pairs.
Pair programming skill deficiencies offer a plausible explanation
of this variance.

Less closely related, there is much pair programming research using
other research methods (such as quantitative surveys and
controlled experiments) and/or working in educational contexts.

Some of these nevertheless describe phenomena similar to those we discuss, 
for example:
a \Co{Knowledge Gap} can lead to a \Co{Power Gap}
\cite{BowJarCul19,DomColHev03},
\Co{Backseat-Driving} is common~\cite{JonFle13},
a \Co{Power Gap} will create problems \cite{DomColHev03}, and
pairs need to learn the skill of pair programming to become
fully productive \cite{VanLas05}.



Our theory shares similarities with Collins' analysis of
\say{Interaction Ritual Chains}~\cite{Collins04}. 
Collins describes emotional energy as 
\say{a feeling of confidence, elation, strength, enthusiasm, and 
	initiative in taking action} 
that leads to confidence whereas low emotional energy can lead to 
depression and withdrawal. 
The latter reminds of our \Co{Disengaging Behavior} concept.
Collins explains that \Co{power rituals} that differentiate between 
those who command and those who follow increase emotional energy in the 
former party and decrease it in the latter. 
\Co{Hierarchical Behavior} can be viewed as such a power ritual. 
Collins also describes \Co{rituals of solidarity} that increase emotional 
energy in all participants. 
\Co{Equalizing Behavior} can be viewed a ritual of solidarity.
Taken together, these similarities suggest that the pair programming behavior patterns
described by our Grounded Theory can also be considered interaction rituals
in the Collins sense.

\section{Conclusions and Further Work} \label{sec_conclusion}

\subsection{Advice for Pair Programming Practitioners}

Whenever you find yourself on the lower end of a \Co{Power Gap},
make meta-discussion a habit (\Co{Equalizing Behavior from-below}),
rather than slipping into
\Co{Defensive Behavior} or even \Co{Disengaging Behavior}.
It will help both the session and your self-esteem.

Whenever you are on the upper end of a \Co{Power Gap},
watch out and avoid accidental \Co{Hierarchical Behavior}.
If you notice symptoms of \Co{Defensive} or \Co{Disengaging Behavior},
apply \Co{Equalizing Behavior from-above} as a counter-measure.
If you find it difficult to identify content suitable for
\Co{Equalizing Behavior}, use meta-discussion instead.

A pair that has improved its pair programming skill enough to reliably
identify and explicate \Co{Power Gaps} as they occur has essentially
solved the problem.
Mini-retrospectives at the end of each session may help getting to this
point~\cite{BocSie20, Shen10, hazmatzo25}.

\subsection{Further Work}

The above advice is currently based on our broader understanding of
pair programming combined with intuition.
Future research should validate and possibly complement these statements.

Quantifying the frequency of the behaviors described in the Grounded Theory
and possibly measuring their impact would be useful steps as well.


%
%

\bibliographystyle{ACM-Reference-Format}
\bibliography{special}

\end{document}